\documentclass[lettersize,journal]{IEEEtran}
\usepackage{amsmath,amsfonts}
\usepackage{algorithmic}
\usepackage{algorithm}
\usepackage{array}
\usepackage[caption=false,font=normalsize,labelfont=sf,textfont=sf]{subfig}
\usepackage{textcomp}
\usepackage{stfloats}
\usepackage{url}
\usepackage{verbatim}
\usepackage{graphicx}
\usepackage{orcidlink}

\usepackage{subcaption} % put this in your preamble
\usepackage[numbers]{natbib}
\usepackage{hyperref}

\begin{document}
\title{Adaptive Substrate Support Based on Thin-Film Piezoelectric Actuators}
\author{Ertuğ Şimşek \textsuperscript{1}\orcidlink{0000-0003-3045-1577},
Bas Jansen\textsuperscript{2},
Marcelo Ackermann\textsuperscript{1}\orcidlink{0000-0003-1191-4265},
Muharrem Bayraktar\textsuperscript{1}\orcidlink{0000-0002-8339-2091}\\

\textsuperscript{1}Industrial Focus Group XUV Optics, MESA+ Institute for Nanotechnology, University of Twente, 7500AE, Enschede, The Netherlands\\
\textsuperscript{2}ASML Netherlands BV, De Run 6501, 5504DR, Veldhoven, The Netherlands
}

        % <-this % stops a space  
        
%\author{
%\IEEEauthorblockN{
%Ertug Simsek,\orcidlink{0000-0003-3045-1577}\textsuperscript{1,2},
%Bas Jansen\textsuperscript{3},
%Marcelo Ackermann,\orcidlink{0000-0003-1191-4265}\textsuperscript{1,2}
%Muharrem Bayraktar, \orcidlink{0000-0002-8339-2091}\textsuperscript{1,2}
%\\}
%\IEEEauthorblockA{\textsuperscript{1}\textit{Industrial Focus Group XUV Optics, University of Twente, The Netherlands\\}} 
%\IEEEauthorblockA{\textsuperscript{2}\textit{MESA+ Institute for Nanotechnology, University of Twente, The Netherlands}\\} 
%\IEEEauthorblockA{\textsuperscript{3}\textit{ASML BV, The Netherlands}\\} 
%\IEEEauthorblockA{Email: e.simsek@utwente.nl}
%}

\markboth{Journal of XX,~Vol.~XX, No.~XX, XX~2026}%
{Simsek \MakeLowercase{\textit{et al.}}: A Sample Article Using IEEEtran.cls for IEEE Journals}

%\IEEEpubid{0000--0000/00\$00.00~\copyright~2026 IEEE}
\maketitle

\begin{abstract}
Advanced lithography scanners require extreme substrate flatness in the range of nanometers to prevent focus errors. 
Such a flatness is challenging to reach and maintain using static substrate supports.  
Active correction concepts using bulk piezoelectric or linear actuators become prohibitive due to wiring and volume limitations, considering the thousands of pillars required. 
In this work, we present to our knowledge the first ever demonstration of an active substrate support concept based on thin-film piezoelectric actuators. 
The working principle, fabrication steps, and characterization of a proof-of-principle device, supported by finite element modeling (FEM), are presented in detail.
The fabricated device shows a displacement above 9 nm, which is sufficient to correct some of the focus errors in lithography systems. We foresee that this new concept may open the way towards smaller chip dimensions and improved yield. 
\end{abstract}


\section{Introduction}
\IEEEPARstart{T}{oday} there is an increasing demand for better performing and more energy efficient computer chips. 
Shrinking the size of integrated circuit features in accordance with Moore's law is one of the most important ways of meeting this demand. 
The minimum feature size is determined by the lithography step. 
Today, the most advanced commercially available lithography systems use light at short-wavelengths to print ever smaller dimensions. %\,\cite{fomenkovHVM_2019}. 
In the lithography process, all of the layers of a chip should be perfectly aligned with the previous layer to have an electrical connection. 
Therefore, advanced lithography machines require very low overlay and focus tolerances down to the sub-nanometer level\,\cite{Bottiglieri_2025_SPIE_OLCD}.

One of the important components of an advanced lithography machine is the substrate support, such as the reticle carrier, wafer carrier, and inspection tools, which provide flatness and hold the substrate during alignment, measurement, and exposure. 
Moreover, the usage of pillar structures on the substrate support enables thermal conditioning by circulating a back-filling gas between them. In addition, these pillars reduce the substrate contact area to prevent the imprints of the substrate backside contamination. 
However, it is challenging to maintain the necessary overlay and focus requirements with the current substrate supports due to effects such as load grid and edge roll-off on the substrates. 
The required local corrections imply a large number of actuators in the range of thousands that have to be steered with sub-nanometer accuracy. 
In the literature, an actuated substrate support concept is proposed to counteract these local imaging errors, but it has not been possible to manufacture it yet\,\cite{steur_2017_AWC,Hurk_2020}. The required accuracy, volume constraints, wire connections, and manufacturing of individual embedded actuators are challenging to reach with available technologies such as bulk piezoelectric or linear actuators.

In this paper, we present to our knowledge the first ever demonstration of active substrate support based on thin-film piezoelectric actuators. 
We present the working principle, fabrication steps, and characterization of a proof-of-principle device with the support of FEM.

\begin{figure*}[!t]
\centering
\includegraphics[width=7.10in]{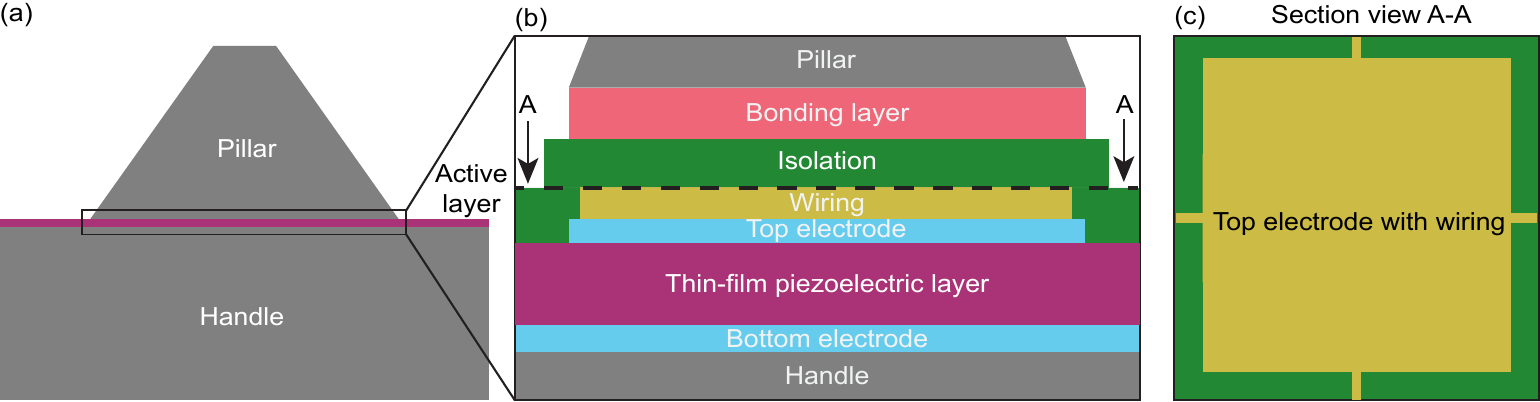}
%(a) \hspace{6cm} (b) \hspace{5.5cm} (c) % labels underneath
\caption{(a) Schematic of a single-pillar active substrate support featuring an integrated piezoelectric actuation layer. (b) Detailed view of the active layer. (c) A section view of the active layer at the top of the wiring layer. (Drawings are not to scale)}
\label{fig: fig_schematic}
\end{figure*}

\section{Principle of Operation and Fabrication}

The presented active substrate support concept consists of a handle and pillar with an integrated active layer between them, as shown in \autoref{fig: fig_schematic}(a). 
Actuating the piezoelectric layer by applying an electric field pushes the pillar in the out-of-plane direction, allowing local correction of focus errors.  

Details of the active layer are shown in \autoref{fig: fig_schematic}(b). 
It is composed of LaNiO\textsubscript{3} (LNO) top and bottom electrodes, a Pb(Zr\textsubscript{0.52}Ti\textsubscript{0.48})O\textsubscript{3} (PZT) layer sandwiched between these electrodes, a Ta/Pt wiring layer to provide electrical connection from outer contact pads to the top electrodes, a silicon nitride (SiN) electrical isolation layer, and a bonding layer from indium/gold (In/Au) to connect the active substrate support and the pillar. 
Top electrodes are wired to four contact pads to have a fail-safe design in case some of the wires have connection issues, as shown in \autoref{fig: fig_schematic}(c).  

The microfabrication process starts with low-stress silicon-rich nitride (SiRN) deposition using low-pressure chemical vapour deposition (LPCVD) on a 525\,\textmu m-thick (100) silicon handle. 
Then, a nanosheet template layer of Ca\textsubscript{2}Nb\textsubscript{3}O\textsubscript{10} (CNO) is deposited using the Langmuir-Blodgett method to allow epitaxial growth of the subsequent bottom electrode, PZT, and top electrode layers\,\cite{bayraktar_2014, Anuj_2015_PZTCNO/Si001}. 
These layers are coated using pulsed laser deposition (PLD) on a 10$\times$10\,mm\textsuperscript{2} substrate, similar to previous work\,\cite{chopra_2017, lucke_2022}. 
After PLD, Ta/Pt metal wires are sputtered and patterned by the lift-off technique. 
The non-covered LNO  is etched using a (1:8) diluted hydrochloric acid (HCl) solution. 
Afterwards, a SiN isolation layer is deposited using plasma-enhanced chemical vapour deposition (PECVD). 
After the isolation layer, a thin gold layer is deposited for bonding and patterned by using the lift-off technique, finalizing the fabrication of the active handle of the substrate support. 

Next, the preparation of the pillar substrate starts with SiRN growth using LPCVD on a 300\,\textmu m-thick (100) silicon substrate. Silicon is selected as the pillar material to make the fabrication process easier using anisotropic Tetramethylammonium hydroxide (TMAH) etching. 
Then, the top surface is patterned with reactive ion etching (RIE) for the hard mask to be used in TMAH etching. 
The bottom surface is coated first by a thin Ta/Pt layer and then by an indium layer using electron-beam evaporation as a preparation for the  bonding step.
The pillar substrate, having a 7$\times$7\,mm\textsuperscript{2} size, is bonded to the active substrate support handle. 
The bonded stack is etched in TMAH for 9~hours until the SiRN layer is reached under the pillar substrate. 
Finally, to reach the contact pads, the SiN layer is removed using RIE. 

\begin{figure}[!t]
\centering
\includegraphics[width=3.5in]{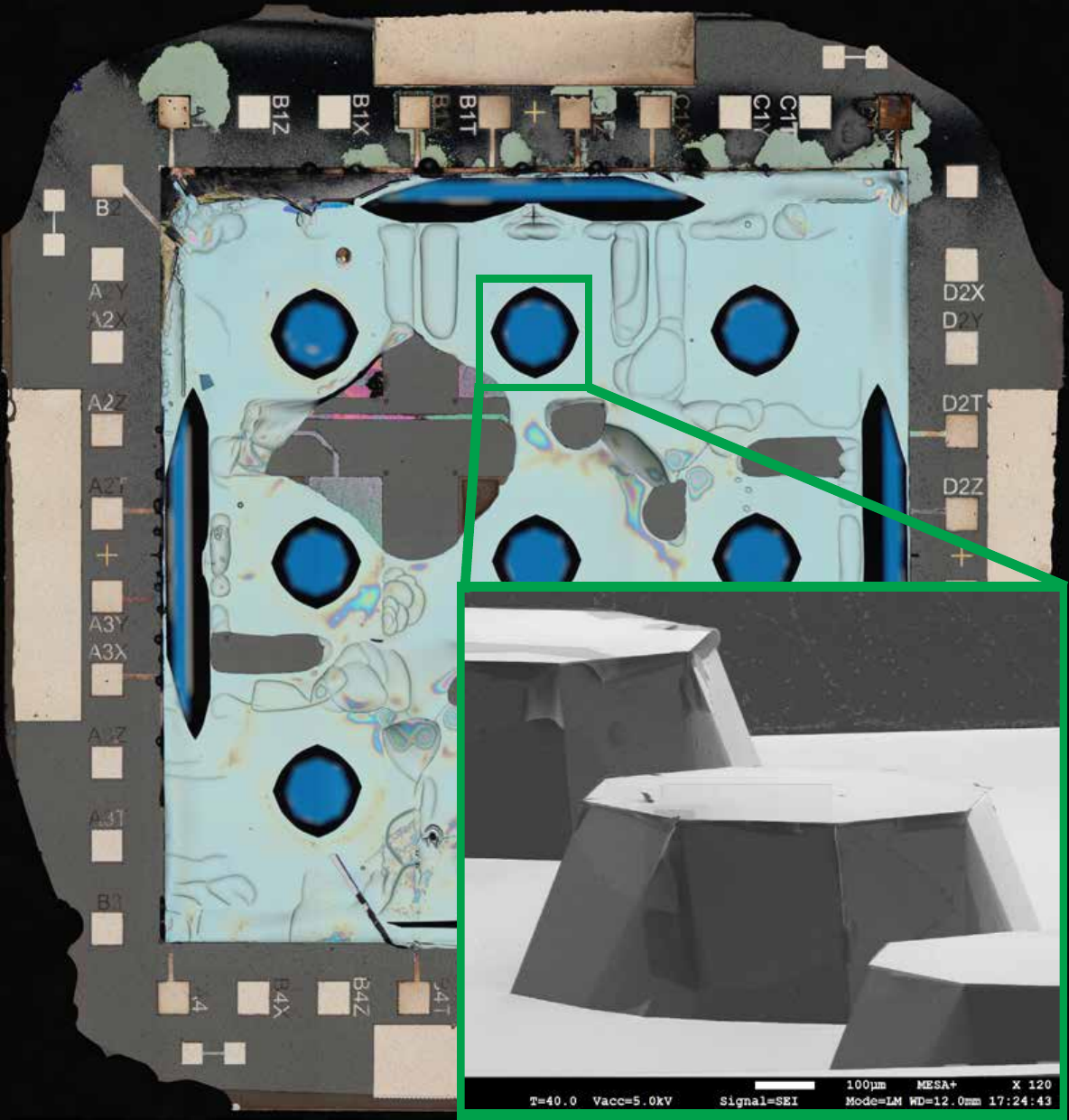}
\caption{Top view of the fabricated device. Inset: SEM image of the pillars.}
\label{fig:fabricated_device}
\end{figure}

The final fabricated device is shown in \autoref{fig:fabricated_device}. 
The contact pads for the bottom (rectangular) and top (square) electrodes are visible at the periphery of the device. A 3$\times$3 pillar matrix is visible in blue in the middle of the device. The inset shows well-defined pyramidal pillars with an octagonal shape resulting from the etch selectivity of different crystal planes of Si.
It is also visible that the corners of the substrate support are also etched in TMAH solution because the handle is made of Si as well. 
Some peeling off of the bonding and isolation layers is observed close to the middle of the handle. 
This can be due to low adhesion of the isolation layer on PZT and long exposure of the SiRN to the TMAH solution. 
Further improvements are possible, but they are beyond the scope of this letter and will be presented in future publications.
All in all, the contact pads and pillars are intact and are ready for measurements.

\section{Results and discussion}
\begin{figure}[!t]
\centering
\includegraphics[width=3.5in]{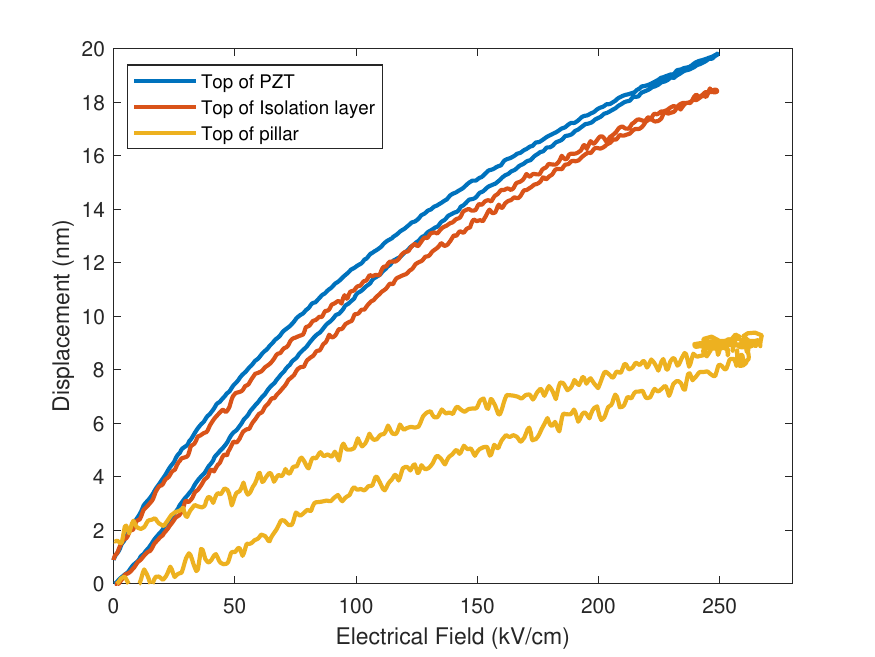}
\caption{Measured displacement at various fabrication steps of the device.}
\label{displacement_field}
\end{figure}

The fabricated device is characterized at various stages of fabrication using a double-beam laser interferometer (DBLI) from aixACCT Systems. For piezoelectric measurements, a triangular signal with a maximum electric field of 250\,kV/cm is applied, unless otherwise stated, to a 4.5\,\textmu m-thick PZT film.
First, the surface displacement is measured after the electrode/piezoelectric thin-film stack is deposited but before the isolation layer. 
The measured displacement is found to be 20\,nm as shown in~\autoref{displacement_field}. 
Secondly, the surface displacement after the isolation layer is measured as 19\,nm. 
Finally, the displacement at the top of the pillar is measured by entering a field of 300\,kV/cm. 
However, due to current leakage in the device, a maximum of 267\,kV/cm could be applied, which is limited by the power supply. 
At the applicable electric field, a displacement of 9.3\,nm is measured. 
Some difference between the measured displacement values is expected, considering the changing boundary conditions of the PZT film and the elastic properties of the added layers. 
For a better understanding of the differences, we model the device in each measurement step using COMSOL Multiphysics 6.4 software. 

% \begin{figure}[!th]
% \centering
% \includegraphics[width=3.5in]{5_Overlay3displacements_2.pdf}
% \caption{Measured displacement at various fabrication steps of the device.}
% \label{displacement_field}
% \end{figure}

To model the device in the first measurement step, we use PZT-5H of the library as the piezoelectric layer.
The electrode layers are not included in the model since their thicknesses are negligible compared to the rest of the layers, but instead the electric field is applied by setting electric boundary conditions to the bottom and top surfaces of the piezoelectric layer. 
The measured surface displacement is mainly determined by d\textsubscript{33}, but in thin films, it is known to be highly dependent on d\textsubscript{31} and geometric parameters such as the ratio of electrode size to substrate thickness\,\cite{Sivaramakrishnan_2018}. 
It should be noted that for a given geometry and elastic parameters the same surface displacement can be reproduced at several combinations of d\textsubscript{33} and d\textsubscript{31} values.
Here, the d\textsubscript{33} of PZT-5H is kept constant considering the previously reported values of similar films\,\cite{d33d31_minh_2017}, and the value of d\textsubscript{31} is swept to reproduce the measured surface displacement. 
The measured surface displacement can be reproduced at a value of d\textsubscript{31}\,=\,278\,pm/V.

In modeling the second measurement, the isolation layer has been added to the model with the material properties from the library. 
%A pre-stress of 250\,MPa is also included in the model according to the previous characterization of similar films. %\,\cite{This Oxfords manual, not a paper}. 
The displacement is found to be 20\,nm from the model, same as in the previous step, while the measured displacement is slightly lower 19\,nm. The values agree within the $\pm$5\% experimental measurement uncertainty of the DBLI. 

In modeling the final measurement, the bonding layer and pillar are added to the model with properties from the library. 
A displacement of 21\,nm is found at the top of the pillar according to the modeling, which is in contrast to the reduction of the measured displacement to 9.3\,nm. 
%Some reduction in displacement compared to the top of the isolation layer is expected due to the low stiffness of the bonding layer. 
%However, this model does not fully explain the measured reduction. 
This reduction may arise due to the high current flow, as observed during the DBLI measurements. 
We identified that there are pinholes in the isolation layer that can in the worst case cause actuation of all nine electrodes instead of actuating only the intended electrode. 
Such an actuation can potentially increase the electrode area, and hence the capacitance to be actuated by nine times, requiring a nine-fold higher current that is limited by the power supply of the DBLI, limiting the full expansion of the PZT film.  
%This can explain the high level of current drawn by the device when it is actuated. 
Such pinholes can be prevented in the future by using an improved isolation layer deposition process or by depositing alternating isolation layers. 
Preventing pinholes and leakage has the potential to increase the displacement at the top of the pillar to 21\,nm according to modeling, and even further by increasing the PZT film thickness. 

\section{Conclusion}
In this work, the first active substrate support based on a thin-film piezoelectric actuator is presented. Using standard microfabrication techniques that are scalable to larger areas, a proof-of-principle device is manufactured and characterized. The device shows a displacement of 9.3\,nm that is sufficient to correct some of the focus errors in advanced chip fabrication. This new design represents a step towards achieving smaller chip dimensions and improving yield by minimizing focus errors.

\section*{Acknowledgments}
This research is funded by the TKI-HTSM (TKI2112P13, WAFER) and ASML.
The authors would like to thank the MESA+ Institute cleanroom staff. 

\bibliography{References_1_Paper2}
\end{document}